\documentclass[conference]{IEEEtran}
\usepackage{cite}
\usepackage{amsmath,amssymb,amsfonts}
\usepackage{algorithm}
\usepackage[noend]{algpseudocode}
\usepackage{comment}

\usepackage{fixltx2e}
\usepackage{textcomp}
\usepackage{booktabs} 
\usepackage[english]{babel}
\usepackage[utf8]{inputenc}
\usepackage{multirow}
\usepackage[normalem]{ulem}
\useunder{\uline}{\ul}{}
\usepackage{graphicx}
\usepackage{caption,subcaption,graphicx}

\newcommand{\etal}{\textit{et al.}}

\begin{document}

\title{Towards Understanding Trends Manipulation in Pakistan Twitter}

\author{
\IEEEauthorblockN{Soufia Kausar, Bilal Tahir, Muhammad Amir Mehmood}
\IEEEauthorblockA{
\{soufia.kausar, bilal.tahir, amir.mehmood\}@kics.edu.pk\\
\textit{Al-Khawarizmi Institute of Computer Science}\\
University of Engineering and Technology, Lahore, Pakistan.
}}
\maketitle

\begin{abstract}
The rapid adoption of online social media platforms has transformed the way of communication and interaction. 
On these platforms, discussions in the form of trending topics provide a glimpse of events happening around the world in real-time. Also, these trends are used for political campaigns, public awareness, and brand promotions. Consequently, these trends are sensitive to manipulation by malicious users who aim to mislead the mass audience. In this article, we identify and study the characteristics of users involved in the manipulation of Twitter trends in Pakistan. We propose `Manipify' -- a framework for automatic detection and analysis of malicious users for Twitter trends. Our framework consists of three distinct modules: i) user classifier, ii) hashtag classifier, and ii) trend analyzer. The user classifier introduces a novel approach to automatically detect manipulators using tweet content and user behaviour features. Also, the module classifies human and bot users.  
Next, the hashtag classifier categorizes trending hashtags into six categories assisting in examining manipulators behaviour across different categories. Finally, the trend analyzer module examines users, hashtags, and tweets for hashtag reach, linguistic features and user behaviour. Our user classifier module achieves $0.91$ accuracy in classifying the manipulators. We further test \textit{Manipify} on the dataset comprising of $665$ trending hashtags with $5.4$ million tweets and $1.9$ million users. The analysis of trends reveals that the trending panel is mostly dominated by political hashtags. In addition, our results show a higher contribution of human accounts in trend manipulation as compared to bots.
Furthermore, we present two case studies of \textit{hashtag-wars} and \textit{anti-state propaganda} to implicate the real-world application of our research.

\end{abstract}


\section{Introduction} \label{sec: intro}

Online social media platforms has emerged as key source of information and socializing during last decade. These platforms strive to maximize user engagement through rapid information dissemination to information savvy users. In this regard, Twitter -- a micro-blogging platform -- provides real-time trends of the most discussed topics in the trending panel~\cite{Twittert40:online}. Due to the extensive reach, such trends have enabled journalists and business analysts to explore breaking news, predict candidate popularity, and product reviews~\cite{Howtocov45:online,karami2018mining,taecharungroj2017starbucks,mccorkle2017using}. In addition, a survey reports that $74$\% of Twitter users utilise this platform as a source of daily news while $34$\% of these users focus on trending topics for this purpose~\cite{rosenstiel2015twitter}. On one hand, Twitter trends are being used to detect breakthrough events, product marketing, and crisis management~\cite{taecharungroj2017starbucks,gencoglu2020causal}. On the other hand, these trends are subjected to manipulation by malicious users to spread false narratives~\cite{assenmacher2020two}.

Recent research reveals that Twitter trends can easily be manipulated by using a small number of automated accounts~\cite{abu2019botcamp}. A new business has emerged where companies are selling Twitter trends ``manipulation as a service". These services use bots and trolls to generate scripted conversations to produce false trends~\cite{Manipula68:online}. A survey reveals that $23$-$27$\% conversations on Twitter related to politics during US elections 2016 are carried out by bot accounts~\cite{howard2016bots}. Also, another research identifies $40$\% bot users disseminating information related to COVID-19 on Twitter~\cite{uyheng2020bots}. Due to such manipulation, critics have also demanded to remove the trending panel as it does not reflect the original trending topics~\cite{Criticsw8:online}.

In general, researchers have focused on examining the trend manipulation by analyzing the activity of human and bot accounts~\cite{zhang2016twitter},\cite{nimmo2019measuring}. Also, the pattern of deletion of tweets related to a trend is studied to detect the possibility of manipulation~\cite{elmas2021ephemeral}. Moreover, a limited number of trending hashtags are manually examined for manipulation~\cite{Dennis2020Manip}. However, these approaches have three major limitations. First, only bot accounts cannot be labelled as manipulators as human accounts are also involved in trend manipulation~\cite{Dennis2020Manip}. Second, manipulators do not necessarily delete tweets after creating a trend. Finally, the manual analysis of constantly emerging new trends is not possible. Moreover, proposed techniques for the identification of spam, bot, fake, compromised, and cloned accounts are not extendable for manipulators due to dissimilarity between their behaviours. To the best of our knowledge, no research has been conducted for the automatic identification of malicious users involved in the manipulation of Twitter trends.

This paper is an effort to study the manipulation of trends in Pakistan. In this regard, we propose a novel framework of `Manipify' to automatically identify and examine the manipulators. The framework consists of three major modules: (i) user classifier, ii) hashtag classifier and (ii) trend analyzer. The first module of the user classifier identifies bots, humans, and manipulators using our developed datasets. Precisely, we introduce a novel method to detect manipulators using content and user behaviour features. Our method achieves an accuracy of $0.91$. Also, the user classifier leverages the profile features of description, URL, followers, friends, status count, and geo-location to identify the bot account with $0.84$ accuracy. In addition, the hashtag classifier categorises hashtags into six classes of i) politics, ii) sports, iii) religious, iv) campaign, v) entertainment, and vi) military using our labelled dataset of 2,384 hashtags. Finally, trend analyzer examines the trends for language distribution of tweets, the reach of a hashtag, and the behaviour of manipulators and bots. To test our framework, we build a PK-Trend dataset containing $665$ trending hashtags, $5.4$ million tweets, and $1.9$ million users from Pakistan. Specifically, we collect trending hashtags and their related tweets for one week in November 2020, December 2020, and January 2021 with a gap of five weeks. 
Our major contributions and key findings are summarized as follows:

\begin{itemize}

\item We introduce a machine learning-based method to automatically detect users involved in the manipulation of Twitter trends. Our approach achieves the accuracy of $0.91$ for manipulator detection.

\item Our analysis trending hashtags from Pakistan reveals that political and campaign hashtags dominate the trending panel with $32$-$42$\% and $16$-$32$\% hashtags, respectively. On the contrary, only $1$-$14$\% hashtags are from the categories such as sports, entertainment, and religion. 

\item We found distinct patterns in user preferences of natural languages with respect to hashtag categories. While the English language is preferred for the entertainment and sports category, Urdu is the frequently used language in tweets related to political and religious hashtags. 

\item Our user analysis highlights that on average our dataset contains $374$K ($52$\%) and $31$K ($4.4$\%) bot and manipulator accounts, respectively. Moreover, $51.5$\% of manipulator accounts are human accounts.

\end{itemize}

The rest of the paper is structured as follows: Section~\ref{sec:relatedwork} presents the related work and Section~\ref{sec:dataset} introduces the developed datasets. In Section~\ref{sec: methodology}, we describe the Manipify framework while its evaluation is presented in Section~\ref{sec:hashscope-eval}. Section~\ref{sec:pktrends_analysis} presents the analysis conducted for Twitter trends in Pakistan using Manipify. Further, we present case studies in Section~\ref{sec:casestudy} to show the real-life applications of our framework. Finally, we conclude our work in Section~\ref{sec:conclusion}.

\section{Related Work} 
\label{sec:relatedwork}

Recently, social media analysis has been adopted to perform topic based sentiment analysis~\cite{reyes2018understanding,alharbi2019twitter},
public opinion-mining~\cite{tavoschi2020twitter}, 
emotion analysis~\cite{lwin2020global},
health surveillance~\cite{honings2021health}, 
crime monitoring~\cite{chung,abbass2020framework},
spam detection~\cite{alom2020deep,madisetty2018neural},
crisis management~\cite{purohit2020ranking,kersten2020happens,lorini2019integrating}, and business marketing~\cite{taecharungroj2017starbucks}. 
In addition, the social media users are examined for malicious user identification~\cite{zhang2016detecting,benevenuto2010detecting,kollanyi2016bots,anwar2020bot}, location inference~\cite{reelfs2019hashtag}, and influencer identification~\cite{gokcce2014twitter}.

\subsection{Twitter trend analysis}

In the last decade, researchers have focused on examining Twitter trends due to their impact on society. For instance, a real-time system was developed for the classification of trending topics into news, current events, memes, and commemoratives~\cite{zubiaga2011classifying}. The authors used features from the tweets text and their metadata for the classification of trends. This categorization of trends was a milestone towards identifying the breaking news and viral memes from Twitter which was helpful for journalists and news agencies. In addition, Zhang~\etal~investigated the possibility of trend manipulation on Twitter~\cite{zhang2016twitter}. Authors experimented with features such as popularity and coverage for Twitter trending topics to inspect features that contribute more towards the prediction of trending. Their analysis also indicated the presence of malicious and spam users manipulating Twitter trends. Similarly, Khan~\etal~proposed a real-time trend detection method by analyzing a stream of tweets~\cite{KHAN2021113990}. They used statistical information retrieval methods to extract important terms. Furthermore, first large scale study on manipulated/fake Twitter trends was conducted by Elmas~\etal~\cite{elmas2021ephemeral}. Authors uncovered the fact that nearly 20\% of the global Twitter trends were a result of manipulation. They also observed that both the bots and compromised accounts were involved in the manipulation of trends. In addition, the authors discovered the attacked keyword reaches the trending panel much faster than the normal trends.

\begin{table*}[t] \centering
\caption{{Description of hashtag categories.}}
\label{tab: cate_description}
\begin{tabular}{|l|l|} \hline
\textbf{Category} & \textbf{Description} \\ \hline \hline
\textbf{Political} &
Hashtags related to political figures, events or slogans.\\ \hline
\textbf{Sports} &
Sports hashtags are those that highlight a sports event, teams or players.\\ \hline
\textbf{Religious} &
The hashtags highlighting religious topics.\\ \hline
\textbf{Campaign} &
All hashtags that are promotional or social campaigns i.e, demanding justice. \\ \hline
\textbf{Entertainment} &
Entertainment hashtags include discussion of media celebrities, music, movies etc. \\ \hline
\textbf{Military} &
The hashtags discussing military events of personnel. \\ \hline
\textbf{Other} &
All hashtags that do not match the description of above mentioned categories.\\ \hline
\end{tabular}
\end{table*}

\subsection{Bot classification}

The identification of bots is a hot topic of research due to their usage and impact on social media content. For example, supervised machine learning techniques have been proposed with user-based features such as the age of the account, user verified, the number of followers and followees~\cite{efthimion2018supervised}. Among the user-based features, the feature of geo-location turned out to be the most informative feature for the identification of bot users. Similarly, Anwar~\etal~used an unsupervised learning approach for bot detection in Twitter~\cite{anwar2020bot}. The data from Canadian elections 2019 was used to perform k-means clustering with the feature-set containing the number of daily tweets, retweet percentage, and daily favourite count. In addition, one-class classification approach was also adopted for bot classification~\cite{rodriguez2020one}. Authors also compared the performance of binary-class and multi-class classification with one-class classification. Also, the real-time systems of BotSlayer~\cite{hui2019botslayer} and BotWalk~\cite{minnich2017botwalk} have been developed for the bot detection. BotWalk used an unsupervised adaptive algorithm with network, content, and user-based features for classification with $0.90$ precision value. Furthermore, Sayyadiharikandeh~\etal~have proposed Ensembles of Specialized Classifiers (ESC) for bot detection~\cite{sayyadiharikandeh2020detection}. The proposed methodology was also deployed in the latest version of Botometer -- an online bot detection tool.

\subsection{Twitter User Analysis}

Twitter users are key to disseminate information and creating trends. Studying the characteristics of these users, Motamedi~\etal~conducted a detailed study on two snapshots of Elite user accounts present on Twitter~\cite{motamedi2020examining}. They investigated features of elite users such as a change in followers, followees, and rank over time. Also, the findings showed that graph relation between elite users formed 14–20 communities. Similarly, analysis on one million Twitter users was carried out to analyze the behaviour of demographic groups~\cite{wood2017does}. The authors' analysis of demographic attributes including gender, ethnicity, and account age of one million Twitter users highlighted that various demographic groups show differences in behaviour. For instance, male users tended to specify their location in their Twitter profile while keeping geo-tagged features disabled. Similarly, the female users preferred to use iPhone or Android devices for Twitter instead of a web browser. In addition, Yaqub~\etal~conducted sentiment analysis two political candidates during 2016 US presidential elections~\cite{yaqub2017analysis}. The analysis showed Trump received more positive sentiment as compared to Hillary Clinton. In addition, the authors analyzed tweets of one million Twitter users to identify their opinion. Their findings revealed that existing opinions were re-shared using the retweet feature instead of building new opinions and arguments.

\subsection{Hashtag classification}
The automatic understanding of hashtags is a challenging task because hashtags are inconsistent and lack standard vocabulary~\cite{gupta2020real}. Previously, the researchers have adopted the labour-intensive and time-consuming path for labelling the trending hashtags. In this regard, Romero~\etal~categorised hashtags into eight pre-defined categories manually to examine the patterns of information dissemination~\cite{romero2011differences}. Jeon~\etal~experimented to build a hashtag recommendation system after their topic classification~\cite{jeon2014hashtag}. The TF-IDF lexical features were extracted from tweets and train the Naive Bayes classifier for hashtag classification. Another algorithm was proposed for the hashtag classification after combining lexical and pragmatic features~\cite{posch2013meaning}. The pragmatic features are related to user profiles such as the number of followers or followees. In addition, open-source content like Wikipedia and Open Directory was utilized for the classification of hashtags~\cite{ferragina2015analyzing}. Such as, Ferragina~\etal~used the Wikipedia graph to devise the Hashtag-Entities (HE) graph which represented the semantic relation between hashtags and their entities~\cite{ferragina2015analyzing}. However, this approach is limited to the hashtags that are available in the Wikipedia graph only.

The literature review revealed that the research community has focused on exploring the possibility of manipulation of Twitter trends but no technique has been presented for the automatic detection of manipulators. In contrast, we propose a machine learning based model for automatic manipulator detection. In addition, we build an analyzer module for a comprehensive analysis of Twitter trends.

\begin{table}[t]\centering
\caption{MT-Dat -- Statistics} \label{tab: dataset_manip}
\begin{tabular}{|l|c|c|c|} \hline
& \textbf{Manipulators} & \textbf{Non-Manipulators} & \textbf{Total} \\ \hline \hline
\textbf{\#Users}  & 510     & 500    & \textbf{1,010} \\ \hline
\textbf{\%Users}  & 50.4\%  & 49.6\% & \textbf{100\%} \\ \hline
\end{tabular}
\end{table}

\begin{table}[t] \centering
\caption{BT-Dat -- Statistics.} \label{tab: dataset_bots}
\resizebox{0.47\textwidth}{!}{
\begin{tabular}{|l|c|c|c|c|c|} \hline
\textbf{Dataset} & \multicolumn{1}{l|}{\textbf{\#Accounts}} & 
\multicolumn{1}{l|}{\textbf{\#Bot}}   & \multicolumn{1}{l|}{\textbf{\%Bot}}   &
\multicolumn{1}{l|}{\textbf{\#Human}} &
\multicolumn{1}{l|}{\textbf{\%Human}} \\ \hline \hline
\textbf{midterm-2018}  & 20,000  & 11,908 & 60\%   & 8,092  & 40\%  \\\hline
\textbf{celebrity-2019}& 4,589   & 0      &  0\%   & 4,589  & 100\% \\\hline 
\textbf{Cresci-2017}   & 11,017  & 7,543  & 68\%   & 3,474  & 32\%  \\\hline \hline
\textbf{BT-Dat} & \textbf{35,606} & \textbf{19,451} & \textbf{55\%} 
& \textbf{16,155} & \textbf{45\%} \\ \hline
\end{tabular}}
\end{table}

\begin{table*}[t] \centering 
\caption{{Ha-Dat -- Statistics.}}
\label{tab:dataset_hashtag}
\begin{tabular}{|l|c|c|c|c|c|c|c|c|} \hline
\textbf{Hashtags}& \textbf{Political} & \textbf{Sports} & \textbf{Religious} & \textbf{Campaign}& \textbf{Entertain} & \textbf{Military} & \textbf{Other} & \textbf{Total} \\ \hline \hline
\textbf{English}& 292 & 120 & 77  & 185 & 117 & 102 & 44 & \textbf{937} \\ \hline
\textbf{Urdu}   & 212 & 18  & 102 & 105 & 35  & 38  & 56 & \textbf{566} \\ \hline
\textbf{English-Urdu} 
&  359 &  89 &  75 &  160 &  75 &  29 &  94 &  \textbf{881} \\ \hline \hline
\textbf{Total} &  \textbf{863} &  \textbf{227} &  \textbf{254} &  \textbf{450}
& \textbf{227} &  \textbf{169} &  \textbf{194} &  \textbf{2,384} \\ \hline
\end{tabular}
\end{table*}

\begin{table*}[t] \centering
\caption{PK-Trends -- Statistics.} \label{tab:casestudydata} 
\begin{tabular}{|l|l|c|c|c|} \hline
\textbf{Sr\# }& \textbf{Dataset} & \textbf{PK-Nov-20} & \textbf{PK-Dec-20} & \textbf{PK-Jan-21} \\ \hline \hline
\textbf{1} & \textbf{Time Period}     & 08 - 15 Nov, 2020 & 13 - 20 Dec, 2020 & 21 - 28 Jan, 2021 \\ \hline
\textbf{2} & \textbf{Unique Trends}   & 1,542     & 1,454     & 1,391   \\ \hline
\textbf{3} & \textbf{Unique Hashtags} & 284       & 188       & 193     \\ \hline
\textbf{4} & \textbf{Local Hashtags}  & 231       & 161       & 141     \\ \hline
\textbf{5} & \textbf{Global Hashtags} & 53        & 27        & 52      \\ \hline
\textbf{6} & \textbf{Unique Keywords} & 1,258     & 1,266     & 1,198   \\ \hline
\textbf{7} & \textbf{Unique Users}    & 1,359,406 & 554,513   & 75,628  \\ \hline
\textbf{8} & \textbf{Unique Tweets}   & 2,990,850 & 2,045,448 & 458,799 \\ \hline
\end{tabular}
\end{table*}

\section{Dataset} \label{sec:dataset}
In this section, first, we focus on building separate datasets to detect manipulators and bots. Next, we describe the process of developing a dataset for the hashtag classification. Finally, we discuss our \textit{PK-Trends} dataset to study Pakistan Twitter trends.

\subsection{Manipulator Detection --  \textit{MT-D\lowercase{at}}}

One needs a gold-standard labelled dataset to train a supervised machine learning model for manipulator's identification. Due to the absence of such a dataset, we develop the \textit{MT-Dat} dataset by manually labelling users as `manipulators' and `organic' users. For the labelling of a user, we observe the features of velocity, volume, and content similarity of tweets posted by a user. Generally, the velocity and volume of tweets related to a topic are considered as key features to determine the trending topic~\cite{HowDoesT15:trending}. Leveraging these features, manipulators post tweets with high volume and velocity to create a fake trend. 
In addition, we notice that manipulators generate a large number of posts with similar content to increase the volume of tweets. Therefore, we also manually observe the similarity of tweets posted while assigning a label. Also, the aim of manipulators is to force a fake trend into a trending panel. Therefore, we only observe the tweets posted by users before the hashtag was first seen in the trending panel for labelling. 
Examining these features, two annotators manually labelled randomly selected $1,010$ users with a mutual agreement of 98\%. In MT-dat, $510$ users were labelled as manipulators and $500$ were labelled as non-manipulators.
Table~\ref{tab: dataset_manip} shows the statics of our dataset.

\subsection{Bot detection dataset -- \textit{BT-D\lowercase{at}}}

Next, we use three publicly available datasets to develop our bot detection dataset (BT-Dat). The first labelled dataset of \textit{midterm-2018} dataset contains information of users and tweets from the US midterm elections 2018~\cite{hua2020characterizing,hua2020towards}. Annotators labelled the user as human if they are actively involved in any political discussions. To label bot accounts, features of tweet time and account creation time are manually observed. The midterm-2018 dataset contains a total of $50,538$ user accounts from which $42,446$ are bot and $8,092$ are human accounts. We only include $11,908$ bots from midterm-2018 in our dataset. Due to the insufficient number of human accounts in midterm-2018, we also use the dataset of \textit{celebrity-2019}~\cite{yang2019arming}. This dataset contains $4,589$ human accounts belonging to prominent public figures. In addition, the third dataset of \textit{Cresci-2017} contains $7,543$ bot and $3,474$ human accounts. The accounts in the dataset are labelled by the Crowdflower contributors~\cite{CrowdFlo75:online}. For human accounts labelling, annotators contacted random Twitter users and ask a question in the natural language. Accounts that answered questions properly are labelled as human accounts. Moreover, Cresci-2017 contains three classes of bots: i) traditional bots, ii) fake followers, and iii) social spambots~\cite{mazza2019rtbust}. We combine users in all three datasets to develop a comprehensive \textit{Bot detection} (BT-Dat) dataset that contains $35,606$ labelled Twitter accounts. Table~\ref{tab: dataset_bots} shows statistics for the four bot detection datasets.

\subsection{Hashtag Classification dataset -- \textit{Ha-D\lowercase{at}}}

As such, the hashtags contain symbols, names of organizations, people, or events joined without using any space~\cite{gupta2020real}. A labelled hashtags dataset is required to understand and classify these hashtags into their respective categories. To this goal, we develop a labelled hashtag dataset by manually annotating trending hashtags into six categories of (i) politics, (ii) sports, (iii) religious, (iv) campaign, (v) entertainment and (vi) military. We collected trending hashtags from Twitter using the Twitter API~\cite{twitterapi/1611467} from July 25, 2018, to August 6, 2021. Using the definitions of hashtag categories given in Table~\ref{tab: cate_description}, two annotators manually labelled randomly selected $2,384$ hashtags. If a hashtag doesn't match the description of any category it is labelled as `other'. We fetch tweets related to these hashtags in English and Urdu language. The hashtags that contain tweets in the English language only are referred to as `English hashtags'. Similarly, hashtags that contain Urdu language tweets only are `Urdu hashtags'. Finally, `English-Urdu hashtags' are those that contain tweets in both languages. The detailed statistics of our dataset are shown in Table~\ref{tab:dataset_hashtag}.

\subsection{PK-T\lowercase{rends} dataset}

\begin{figure}[t] \centering
\includegraphics[width=8cm,keepaspectratio]{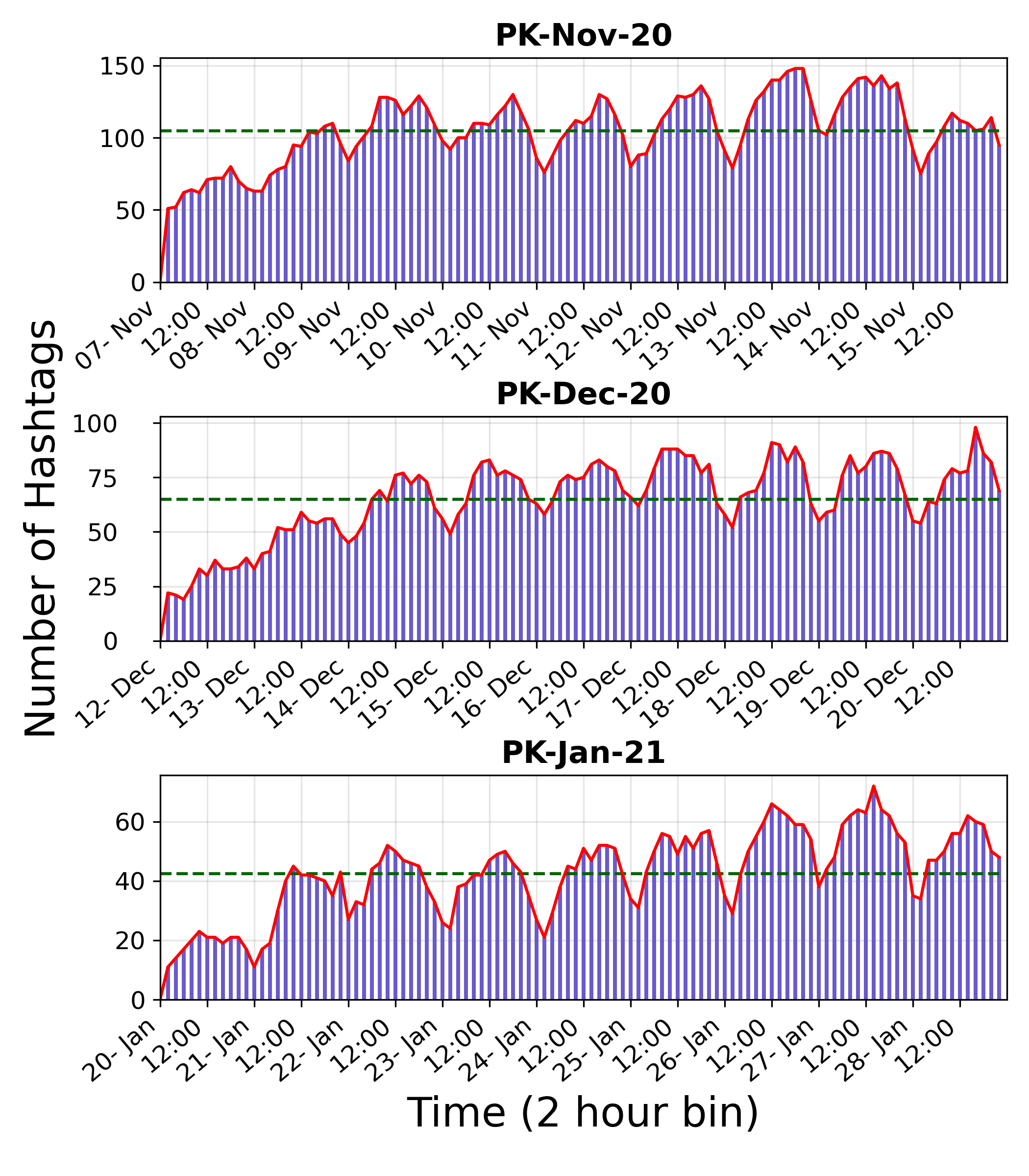}
\caption{Number of hashtags -- PK-Trends dataset.} \label{fig:tagrate}
\end{figure}

Finally, we discuss \textit{PK-Trends} dataset collected to study different aspects of trending hashtags in Pakistan. For this purpose, first, we collect trending hashtags in Pakistan from the online service of GetDayTrends~\cite{Pakistan66:online} which provides the list of trending hashtags on an hourly basis. We create three datasets by fetching trending hashtags for one week in November 2020 (\textit{PK-Nov-20}) , December 2020 (\textit{PK-Dec-20}), and January 2021 (\textit{PK-Jan-21}). We deliberately take these samples with a gap of five weeks to explore trending hashtags dynamics. In addition, we define a time window to fetch tweets related to the trending hashtags. This time window includes tweets from one day before to one day after the hashtag is seen in the trending panel. Moreover, the Python language library of Twint~\cite{twint} is used to fetch the tweets for a three-day time window. We note that Twint could not scrap ``retweets''. Therefore, cumulatively, \textit{PK-Trends} dataset contains $5.4$ million ``original'' tweets posted by $1.9$ million users. Furthermore, PK-Nov-20, PK-Dec-20, and PK-Jan-21 contain $284$, $188$, and $193$ unique trending hashtags, respectively. Table~\ref{tab:casestudydata} shows statistics of PK-Trends dataset. Figure~\ref{fig:tagrate} shows the number of trending hashtags containing related tweets in PK-Trends for seven days with 2 hours bins for all datasets. We observe a periodic pattern in the number of unique trending hashtags with a peak around midday (11 AM-3 PM) consistent with prior studies~\cite{TwA:online}. This observation highlights that users discuss more unique hashtags during daytime. On average, PK-Nov-20, PK-Dec-20, and PK-Jan-21 contain tweets related to $104$ ($36.6$\%), $64$ ($34.1$\%), and $42$ ($21.7$\%) trending hashtags in each bin, respectively. Also, the PK-Trends dataset contains tweets posted in multiple languages. To study the natural language distribution, we use the meta-information of tweet language provided by Twitter. We found that English is the most frequent language used in PK-Trends dataset with $3.2$ million tweets. Whereas, only $0.48$ million tweets are posted in the Urdu language. It has $0.67$ million tweets marked as unknown language and $1.1$ million tweets from ``other'' languages.

\section{Manipify -- Framework} \label{sec: methodology}

\begin{figure*}[t] \centering
\includegraphics[width=0.95\textwidth, keepaspectratio]{{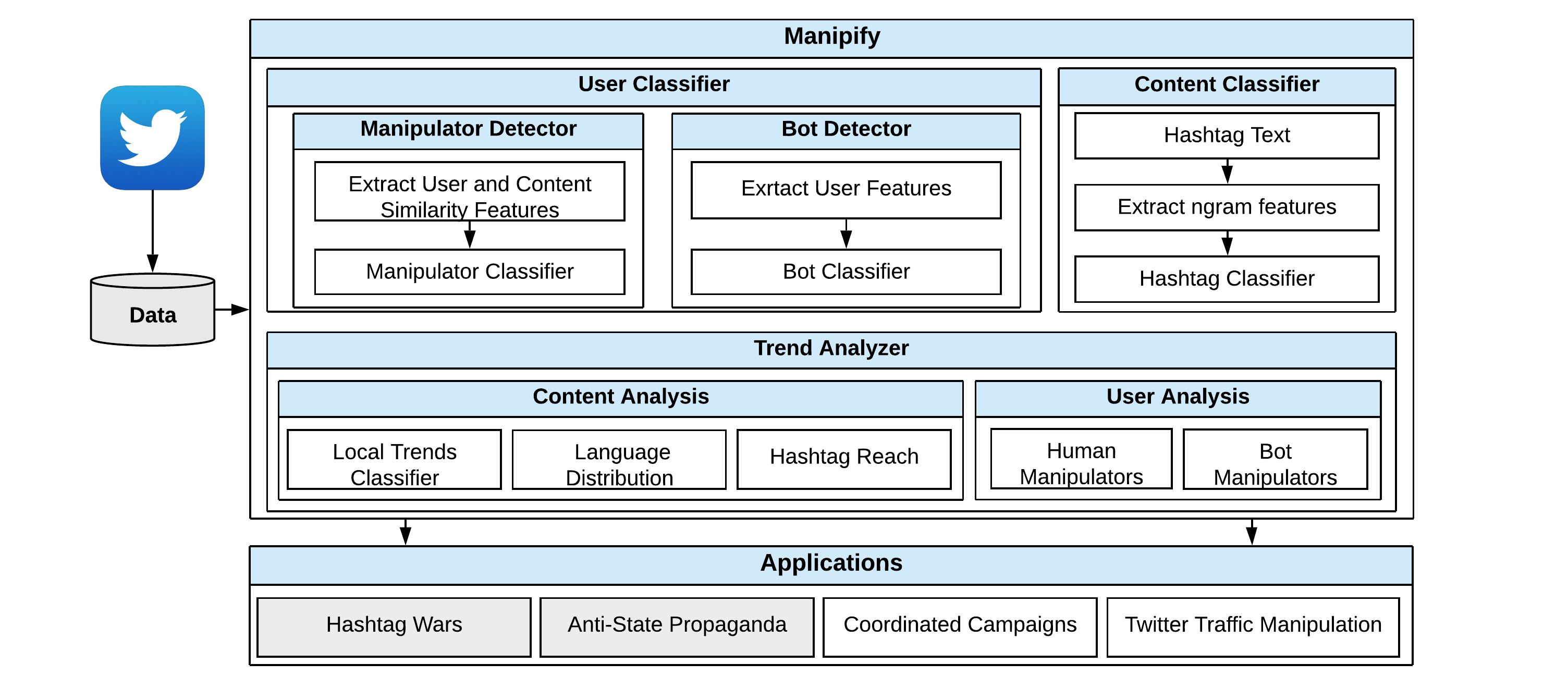}}
\caption{Research methodology -- Flow diagram} \label{fig: architecture}
\end{figure*}

In this section, we describe the architecture of our proposed Manipify framework. First, we explain the user classification modules. Next, we discuss the methodology of hashtag classification. Finally, we present the trend analyzer module proposed to understand the various dynamics of users.

\subsection{Manipulator Detection}

The classification of manipulators requires distinct features related to users to train the machine learning model. In literature, there is no study available which automatically identifies the manipulators using such features. Hence, we design five features related to users which are : 1) number of total tweets by a user ($Tweets$), 2) number of tweets before trend time ($Tweets$\textsubscript{before}), 3) average time between consecutive tweets after trend time ($Time$\textsubscript{after}), 4) average time between consecutive tweets before trend time ($Time$\textsubscript{before}), and 5) content similarity score ($Sim$\textsubscript{score}).
As such, the volume and velocity of tweets containing a hashtag are key factors to determine the trending hashtags~\cite{HowDoesT15:trending}. 
In literature, the manual analysis of manipulators reveals that they post a large number of tweets ($Tweets$) using a hashtag to create trend ~\cite{Dennis2020Manip}. Particularly, it is observed that these users post tweets before trend time ($Tweets$\textsubscript{before}) as `organic user' generally use the hashtag after it is seen in the trending panel. In addition, the velocity of tweets containing a hashtag is the key factor to determine the trend. Therefore, we consider the average time between tweets posted by user before ($Time$\textsubscript{before}) and after ($Time$\textsubscript{after}) trending time of a hashtag. We calculate the velocity of tweets before and after trending time as we believe manipulators limit the activity after trending which impacts the overall value of tweets velocity. 

Finally, we calculate the similarity score ($Sim$\textsubscript{score}) of tweets posted by a user with the aim of identifying the manipulators posting similar tweets to increase the volume. In addition, posting and deleting a large number of tweets using the same hashtag is a violation of Twitter's platform policy~\cite{TwitterManip}. To calculate the content similarity score, we make use of natural language processing techniques. The similarity score of all tweets posted by users related to a hashtag is computed by using the overlapping n-grams in tweets. This is worth noting that the comparison of all possible n-grams of tweets with length greater than one is done for score calculation. Lets a user ($U$) post $n$ number of tweets related to a hashtag ($H$). We calculate the similarity score ($Sim$\textsubscript{score}) of users concerning hashtags by computing the ratio of overlapping n-grams and total n-gram of all tweets posted by users using Equation~\ref{eq:manip}.

\begin{equation} \label{eq:manip} \centering
 Sim_{score}(U) = \dfrac{\sum_{w=2}^{n}(ngram_{W} * freq(ngram))}{\#tweets}
\end{equation}

It should be noted that we only utilise the original tweets of a user for calculating the classification features and retweets are not considered. Utilising these five features, we train a Logistic Regression classifier to classify the manipulators.

\subsection{Bot Detection}
Finally, we use the user behaviour and activity features for bot classification. We extract nine binary features related to user account~\cite{efthimion2018supervised}. They are: 1) User description, 2) URL exists in description, 3) Friend count$>$1000, 4) Follower count$<$30, 5) User geo-location, 6) List count$>$0, 7) Statuses count$>$0, 8) URL exists in profile, and 9) User verified account. In addition to these, we also use the features of followers count and friends count. The presence of description highlights the probability of human users because bot accounts lack such customized information. Similarly, human users tend to have a lower number of friends and the user with a large friend count are considered bots. In addition, the bot account generally does not enable geo-location for interaction. Moreover, the verification of accounts by Twitter is an indication of human accounts. These verified accounts belong to celebrities, organizations, or political parties. We refer to the bot classifier as ``BotCat'' in the rest of the paper. We train and evaluate the performance of the Logistic Regression classifier with the split ratio of 70:30 for bot detection.

\subsection{Hashtag Classification}

Figure~\ref{fig: architecture} shows the architecture diagram of Manipify. To build a hashtag classification module, we leverage the lexical features of tweets for the classification of hashtags as done by of~Jeon~\etal~\cite{jeon2014hashtag}. 
First, during the pre-processing of tweet text, we remove all symbols, hashtags and URLs. Next, we extract 1,3-ngrams utilising the statistical technique of Term Frequency Inverse Document Frequency (TF-IDF) from all pre-processed tweets related to a hashtag. Using the extracted TF-IDF features, we train a Logistic Regression classifier with default parameters for classification. It is crucial to mention here that we train separate TF-IDF vectorizers and models for English and Urdu hashtags. Moreover, we train a series of binary classifiers for each hashtag category using the `one-vs-all' approach due to its better performance compared to multi-class classifiers~\cite{perez2020object,takenouchi2018binary}. We assign the category of the classifier with the highest probability to the hashtag. The hashtag is assigned the label of `Other' if binary classifiers of all categories give the probability less than $0.5$. Moreover, we need to calculate the minimum number of required tweets to accurately classify the topic of a hashtag. In this regard, we use 100 as a minimum number of tweets required to classify the hashtag in accordance with literature~\cite{jeon2014hashtag,lee2011twitter}.

\begin{table*}[t]  \centering
\caption{{Hashtag classification results on HT-Dat dataset.}} \label{tab:result_multiclass}
\begin{tabular}{|l|c|c|c|c|c|c|c|c|}  \hline
\textbf{Dataset} & \textbf{Measure} & \textbf{Political} & \textbf{Sports} & \textbf{Religious} & \textbf{Campaign} & \textbf{Entertainment} & \textbf{Military} & \textbf{Other} \\ \hline \hline
\multirow{3}{*}{\textbf{English}}
&\textbf{Precision} & 0.82 & 0.93 & 0.96 & 0.78 & 0.90 & \textbf{0.97} & 0.78 \\ \cline{2-9} 
& \textbf{Recall}   & 0.82 & 0.92 & 0.92 & 0.72 & 0.88 & \textbf{0.97} & 0.71 \\ \cline{2-9} 
&\textbf{F1-score}  & 0.82 & 0.93 & 0.94 & 0.76 & 0.89 & \textbf{0.97} & 0.66 \\ \hline \hline
\multirow{3}{*}{\textbf{Urdu}}   
& \textbf{Precision}& 0.83 & 0.91 & \textbf{0.95} & 0.78 & 0.80 & 0.89 & 0.63 \\ \cline{2-9} 
& \textbf{Recall}   & 0.82 & 0.83 & \textbf{0.92} & 0.72 & 0.65 & 0.79 & 0.61 \\ \cline{2-9} 
& \textbf{F1-score} & 0.82 & 0.85 & \textbf{0.93} & 0.74 & 0.65 & 0.81 & 0.58 \\ \hline \hline
\multirow{3}{*}{\begin{tabular}[c]{@{}c@{}}\textbf{English} \\ \textbf{- Urdu}\end{tabular}}
&\textbf{Precision} & 0.88 & 0.92 & \textbf{0.94} & 0.85 & 0.76 & 0.83 & 0.38 \\ \cline{2-9} 
& \textbf{Recall}   & 0.91 & \textbf{0.89} & 0.85 & 0.72 & 0.83 & 0.86 & 0.44 \\ \cline{2-9} 
&\textbf{F1-score}  & 0.89 & \textbf{0.90} & 0.89 & 0.78 & 0.79 & 0.84 & 0.41 \\ \hline
\end{tabular} \end{table*}

\begin{table*}[t] \centering
\caption{Sample of local and global trending hashtags.} \label{tb:local}
\begin{tabular}{|l|l|l|l|l|l|l|l|} \hline
\textbf{Sr\#} & \textbf{Hashtag} & \multicolumn{2}{l|}{\textbf{Trending Time}}
& \multicolumn{3}{l|}{\textbf{Features}} & \textbf{Local} \\ \hline 
\textbf{} & \textbf{} & \textbf{Date} & \textbf{Hours} & \textbf{1st trend} &
\textbf{\begin{tabular}[c]{@{}l@{}}\# \\ \end{tabular}} & \textbf{\begin{tabular}[c]{@{}l@{}}World\\ \end{tabular}} & \textbf{} \\ \hline \hline
\textbf{1} & \#SamsungPakistan & 24-01-2021 & 14:00-15:00 & Pakistan & 0 & False & True \\ \hline
\textbf{2} & \#Fajr & 26-01-2021 & 02:00-05:00 & Pakistan & 1 & False & True \\ \hline
\textbf{3} & \#MotivationalQuotes & 25-01-2021 & 02:00-05:00 & Pakistan & 2 & False & True \\ \hline\hline
\textbf{4} & \#POTUS & 21-01-2021 & 00:00-04:00 & Australia & 12 & False & False \\ \hline
\textbf{5} & \#UFC257 & 24-01-2021 & 02:00-12:00 & Japan & 58 & True & False \\ \hline
\textbf{6} & \#T10League & 28-01-2021 & 16:00-17:00 & India & 2 & False & False \\ \hline
\end{tabular} 
\end{table*}

\subsection{Trend Analyzer}
Next, we turn our attention to the analyzer module which is designed to perform an in-depth analysis of trending hashtags and users. First, the module focuses on analyzing the content of hashtags by calculating the distribution of local trends, natural language, and hashtag reach. 
Twitter trending panel often consists of hashtags related to \textit{local} and \textit{global} topics. Hashtags related to global topics like \textit{\#MUFC} and \textit{\#CristianoRonaldo} are discussed all around the world and these trends are not specifically related to local topics. We limit the scope of the Manipify to study different aspects of local trending hashtags due to two major observations. First, global hashtags contain tweets in multiple languages. Second, these hashtags are adopted internationally and the context of hashtags varies with the cultural context of users~\cite{sheldon2020culture}. These features of global hashtags made the understanding and classification of these hashtags a challenging task. Therefore, we propose a classification model of PK-Hash-local to identify local trends for analysis. For classification, the following meta-information related to hashtag is fetched: i) Trending time of hashtag in the target country, ii) Trending time of hashtag in other countries, iii) List of countries hashtag is seen in trending panel, and iv) Whether Hashtag has trended worldwide. We use this meta-information of hashtags to derive three features for the classification. The feature of ``1st trend'' is used to determine the country where a hashtag was first trended. The hashtag is likely to be related to a local topic if it is first seen in the trending panel of the target country. Similarly, the feature of ``number of countries trended'' counts the number of locations where the hashtag is seen in the trending panel. With the higher number of this feature, the hashtag has less probability of being a local trend. Finally, the self-descriptive binary feature of ``Worldwide trended'' shows that the hashtag is global or local. Table~\ref{tb:local} shows sample labelled hashtags with features from PK-Trends. Finally, we use meta-information features of hashtags to train the decision tree classifier.

To study the natural language of tweets related to a hashtag, we utilise the meta-information of tweet language provided by Twitter to calculate the language distribution as described in Section~\ref{sec:dataset}. Finally, the analyzer calculates the reach of a hashtag which is a measure to calculate the number of users who potentially have viewed the hashtag~\cite{tweetbinder}. It also gives an approximation of the extent to which a trending hashtag can affect the Twitter community. For instance, the hashtag used in the tweet of a celebrity with millions of followers has a greater effect on the Twitter community as compared to a regular user. The possible reach of a hashtag is computed by summing the number of followers of all users who tweeted the hashtags. To calculate the reach ($Reach$) of a hashtag ($H$) tweeted by unique users ($n$), the Equation~\ref{eq:reach} is used.

\begin{equation} \label{eq:reach} \centering
 Reach_{H} = n+ {\sum\limits_{i=1}^{n}{Followers_{i}}}
\end{equation}

Furthermore, the trend analyzer performs the analysis of the user by studying the behaviour of bots and manipulating users. For user analysis, first, we detect manipulators and bots to investigate their characteristics. Next, their distribution is analysed in different hashtag categories. Besides, a time series analysis is presented for a sample of hashtags to scrutinize the behaviour of bots, humans, manipulators and organic users.

\begin{table}[t] \centering
\caption{{Classification performance of manipulator and bot detection}}
\label{tab:bot_results}
\begin{tabular}{|c|l|c|c|c|c|} \hline
\textbf{Sr.} & \textbf{Class} & \textbf{Precision} & \textbf{Recall} & \textbf{F1-score} & \textbf{Accuracy} \\ \hline \hline
\textbf{1} & \textbf{Manipulators} & 0.90 & 0.92 & 0.91 & 0.91 \\ \hline
\textbf{2} & \textbf{Organic}      & 0.92 & 0.89 & 0.90 & 0.91 \\ \hline \hline
\textbf{} & \textbf{Total}  & \textbf{0.91} & \textbf{0.91} & \textbf{0.91} & \textbf{0.91} \\\hline\hline
\textbf{3} & \textbf{Human} & 0.79 & 0.82 & 0.81 & 0.86\\ \hline
\textbf{4} & \textbf{Bots}  & 0.89 & 0.88 & 0.89 & 0.86\\ \hline \hline
\textbf{} & \textbf{Total}  & \textbf{0.84} & \textbf{0.85} & \textbf{0.85} & \textbf{0.86} \\\hline
\end{tabular} \end{table}

\begin{figure*}[t] \centering
\includegraphics[width=0.85\textwidth, keepaspectratio]{{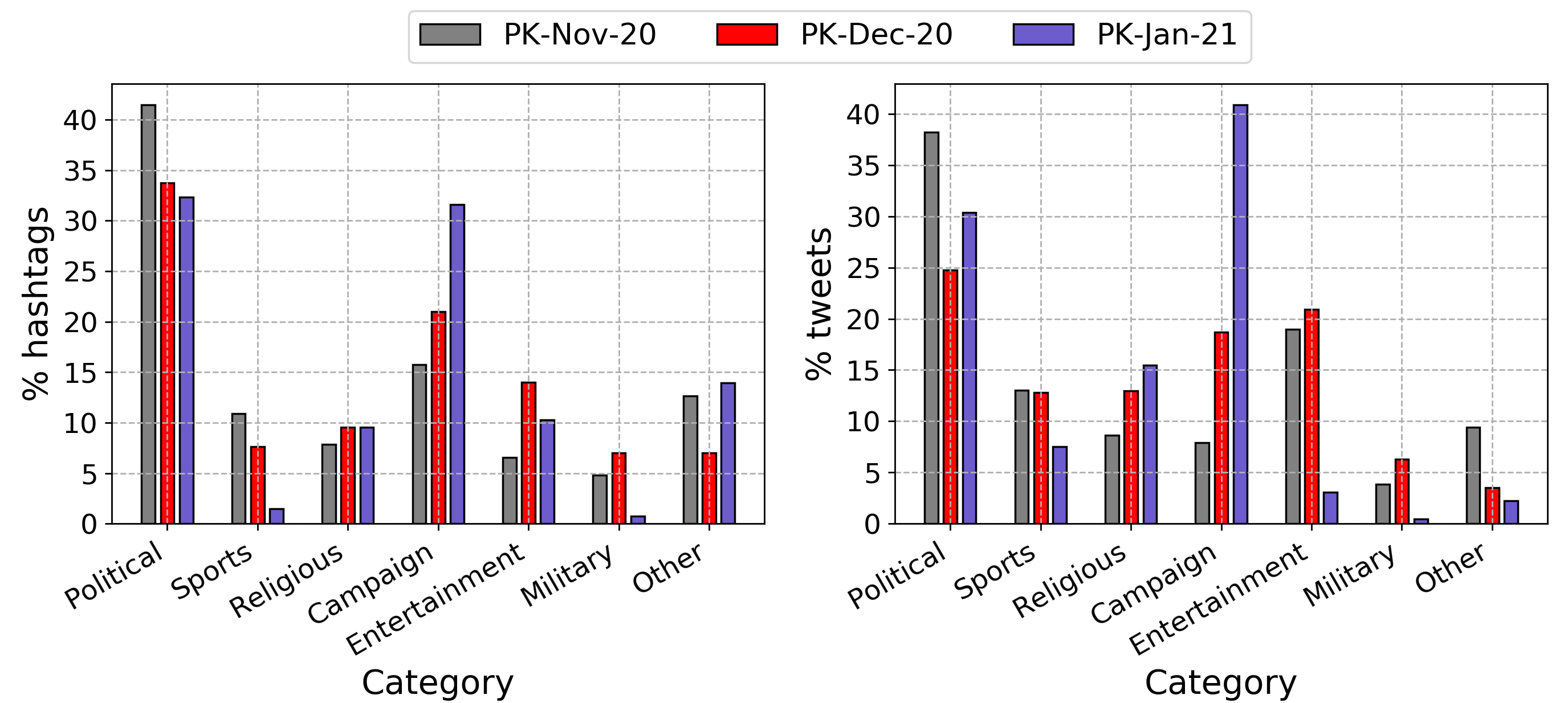}}
\caption{Percentage of hashtags for each category -- PK-Trends-Local dataset.} \label{fig:catdist}
\end{figure*}

\section{Manipify Evaluation} \label{sec:hashscope-eval}

In this section, first, we describe the performances of manipulators and bot classifiers. Next, we dive in to the detailed results of hashtag classification. Finally, we discuss the PK-Hash-local classifier used to create a dataset for hashtags related to Pakistan only.

\subsection{Manipulator Detection}

Table~\ref{tab:bot_results} presents the results of manipulator detection. The classifier achieves an overall accuracy and F1-score of $0.91$ each. However, the manipulators are classified with a higher F1-score of $0.91$ as compared to the organic users. The examination of weights assigned to features used for classification reveals the highest contribution of the feature of \textit{Tweets\textsubscript{before}}. This feature is evidence of manipulation as our manual analysis proves that manipulators are more active before the trend time of a hashtag. In addition, the second-highest weight is assigned to the \textit{Sim\textsubscript{score}} feature as the large similarity between tweets of a user shows manipulative behaviour~\cite{TwitterManip}. On the other hand, the most informative features to classify organic user is \textit{Time\textsubscript{after}} and \textit{Time\textsubscript{before}}. The manipulators tend to post tweets with the intent to increase the volume and velocity as these are the key factors for determining a trend~\cite{HowDoesT15:trending}. Whereas, the organic users do not have any pattern in the average time between their consecutive tweets.

\subsection{BotCat}

Table~\ref{tab:bot_results} shows the classification performance of our BotCat with BT-Dat dataset. We note that the bot classifier achieves $0.86$ accuracy on the BT-Dat dataset with precision, recall, and F1-score of $0.84$, $0.85$ and $0.85$, respectively. The $0.89$ precision for bot class compared to $0.79$ precision value for human class indicates that the classifier identifies bot accounts more accurately as compared to human accounts. To investigate the reason for the lower performance of the BotCat for human accounts, we analyze the weights assigned by the trained classifier to the classification features. We observe that the classifier assigns the highest weight to feature of \textit{follower count $<$30} to bot class explaining that users with follower count less than 30 are more likely bot account. Similarly, the feature of \textit{user verified} is assigned the highest weights for human accounts. In addition, we analyze bot accounts and observe that the few bot accounts have a fake name, profile picture, and the given description. Moreover, these accounts replicate the behaviour of human users which results in mis-classification of these users as human accounts~\cite{cresci2017paradigm}.

\subsection{Hashtag Classification}

We train and evaluate the hashtag classifiers for English and Urdu hashtags separately. 
Table~\ref{tab:result_multiclass} gives the detailed classification results. First, we discuss the performance of the classifier for English hashtags. The English hashtag classifier achieved an accuracy of $0.84$ with a $0.70$ F1-score. However, we observe the highest F1-score of $0.97$ for the \textit{military} hashtags. Similarly, the \textit{sports} and \textit{religious} hashtags have comparable F1-scores of $0.93$ and $0.94$. Whereas, the \textit{Other} category hashtags achieve the lowest F1-score of $0.66$. Finally, the \textit{political}, \textit{campaign} and \textit{entertainment} hashtags have $0.82$, $0.76$ and $0.89$ F1-scores. The low F1-score of hashtags for some categories is due to the high lexical diversity in samples of these classes. In addition, the number of training samples also affect the classification performance of binary classifiers.

Next, we divert our attention towards the classification results of Urdu hashtags. The Urdu hashtag classifier attains $0.79$ accuracy and $0.63$ F1-score. Here, the \textit{religious} hashtags are classified with highest F1-score of $0.93$. Similar to the English hashtags, the \textit{Other} hashtags are classified with the lowest F1-score of $0.58$. The \textit{entertainment} hashtags also achieve a low F1-score of $0.65$. Moreover, the categories of \textit{political}, \textit{sports}, \textit{campaign} and \textit{military} attain F1-scores of $0.82$, $0.85$, $0.74$ and $0.81$. Overall, we observe lower F1-scores for the Urdu language binary classifiers as compared to the English classifiers. We attribute this result to the lower number of samples in the training data for Urdu hashtags. 

The classification of English-Urdu hashtags with our framework presents a interesting challenge that which classifier (English or Urdu) should be used for such hashtags. In this regard, first, we classify the hashtag with both English and Urdu classifiers using tweets of respective languages. Next, we compare the probabilities for each class assigned by both classifiers. Finally, we assign the label of the category with the highest probability assigned by either classifier. Using this approach, we observe the English-Urdu classifier achieves $0.84$ accuracy and $0.79$ F1-score. However, we notice the best performance for the \textit{sports} hashtags with $0.87$ F1-score. The \textit{religious} and \textit{political} category hashtags have comparable performance with $0.89$ F1-score each. Whereas, the \textit{campaign}, \textit{entertainment} and \textit{military} hashtags are classified with $0.78$, $0.79$ and $0.84$ F1-scores respectively. Consequently, the \textit{Other} category hashtags are classified with lowest F1-score of $0.41$. We notice that this approach achieves better performance compared to the Urdu language classifier.

\subsection{PK-Trends-Local}

Leveraging the PK-Hash-local classifier, we create the PK-Trends-Local dataset containing the trends related to Pakistan. In order to classify the `global' and `local' trends, first, we create the labelled dataset by manually labelling the $193$ hashtags of the PK-Jan-21 dataset into the local and global category. With manual classification, $141$ ($73$\%) hashtags are labelled as local while $52$ ($27$\%) are labelled as global hashtags. Next, the meta-information related to hashtags in the PK-Trends dataset is fetched from GetDayTrends. Using the hashtag features (explained in Section~\ref{sec: methodology}) of the labelled dataset and the split ratio of 70:30, the classifier achieves the accuracy of $0.97$. Next, we classify trends of PK-Nov-20 and PK-Dec-20 using the trained classifier and identify $231$ and $161$ local hashtags in PK-Nov-20 and PK-Dec-20, respectively. Table~\ref{tab:casestudydata} shows the distribution of local and global trends in PK-Trends.

\begin{figure*}[t] \centering
\includegraphics[width=0.8\textwidth]{{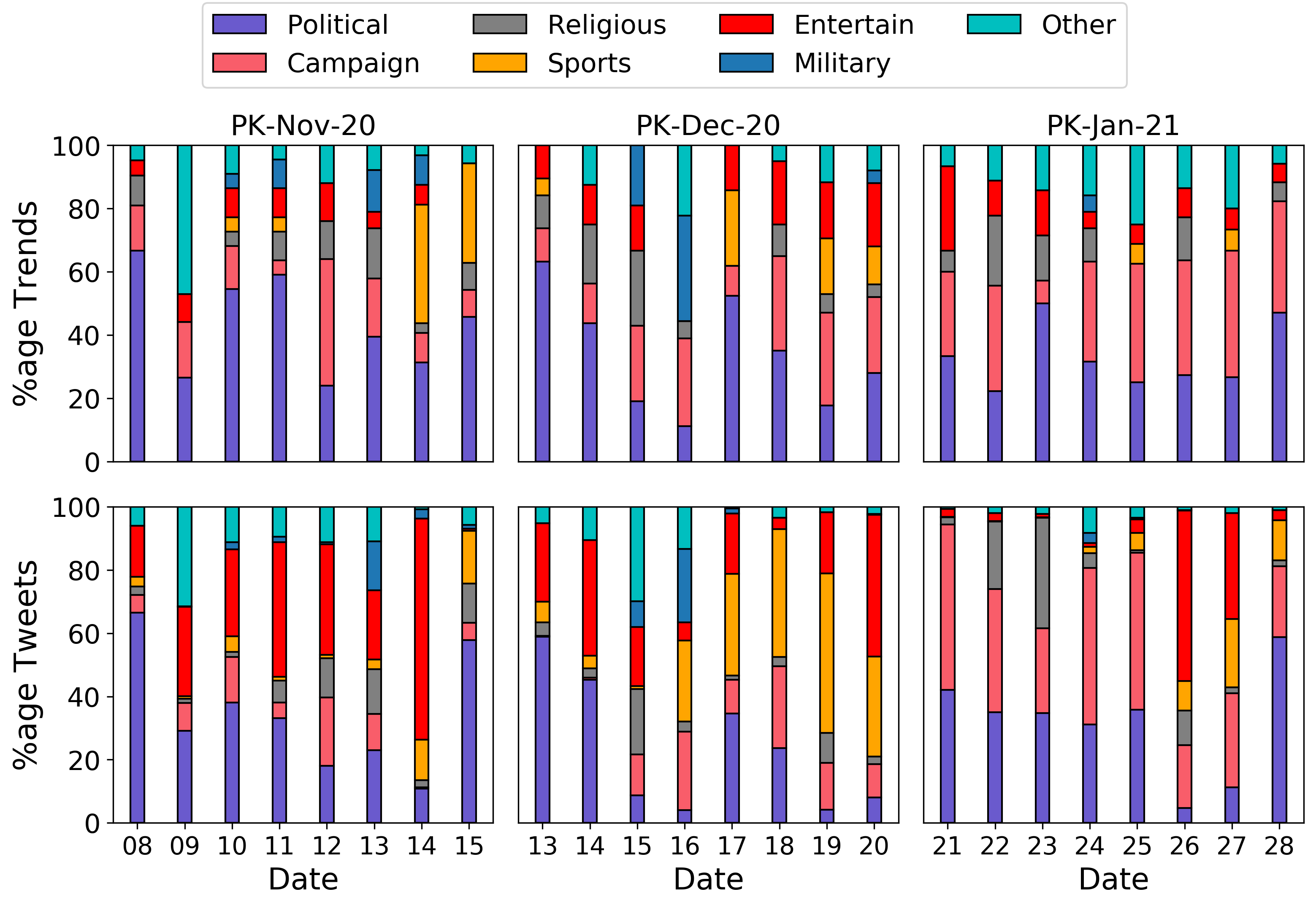}}
\caption{{Per day distribution of trends and tweets for each category -- PK-Trends-Local dataset.}} \label{fig:trends}
\end{figure*}

\begin{figure*}[t] \centering
 \begin{subfigure}{0.5\textwidth} \centering
 \includegraphics[width=8cm,keepaspectratio]{{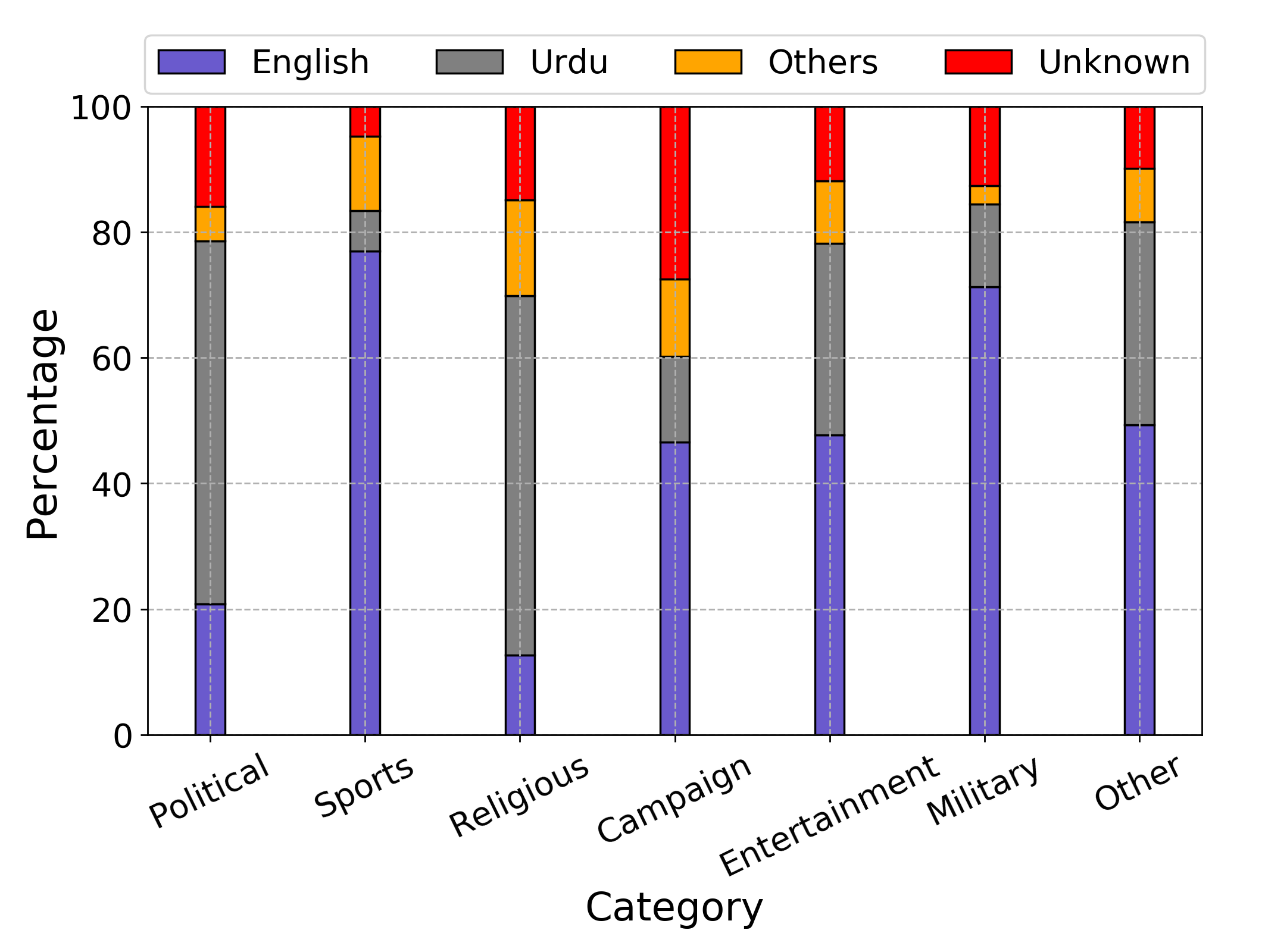}}
 \caption{Category-wise language distribution of tweets -- PK-Nov-20 dataset.} \label{fig:lang}
 \end{subfigure}%
 ~ 
 \begin{subfigure}{0.5\textwidth} \centering
 \includegraphics[width=8cm,keepaspectratio]{{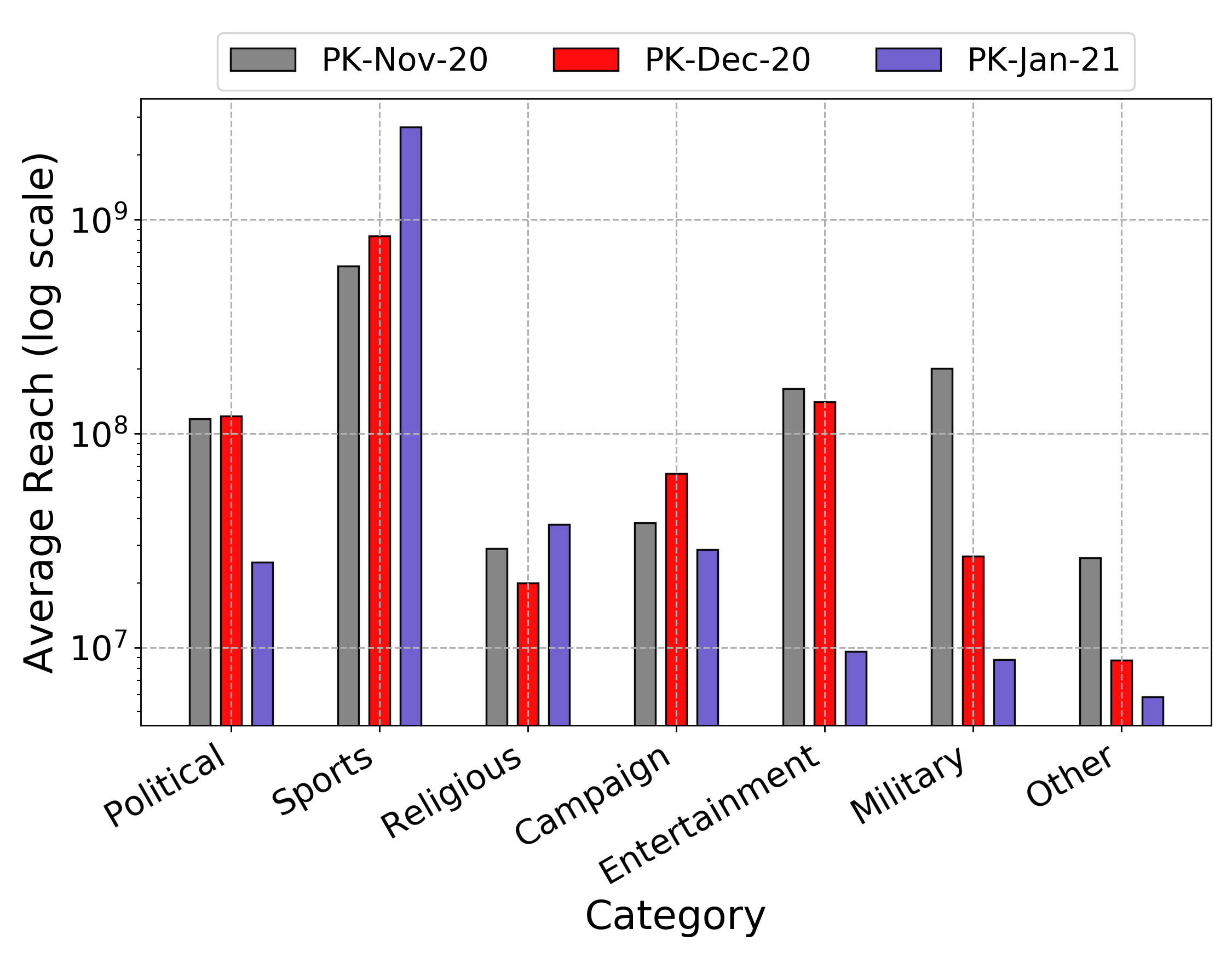}}
 \caption{Average reach of trends for each category -- PK-Trends-Local dataset.} \label{fig:reach}
 \end{subfigure}
 \caption{Category distribution according to language and average reach.}
\end{figure*}

\begin{figure}[t] \centering
\includegraphics[width=8.6cm,keepaspectratio]{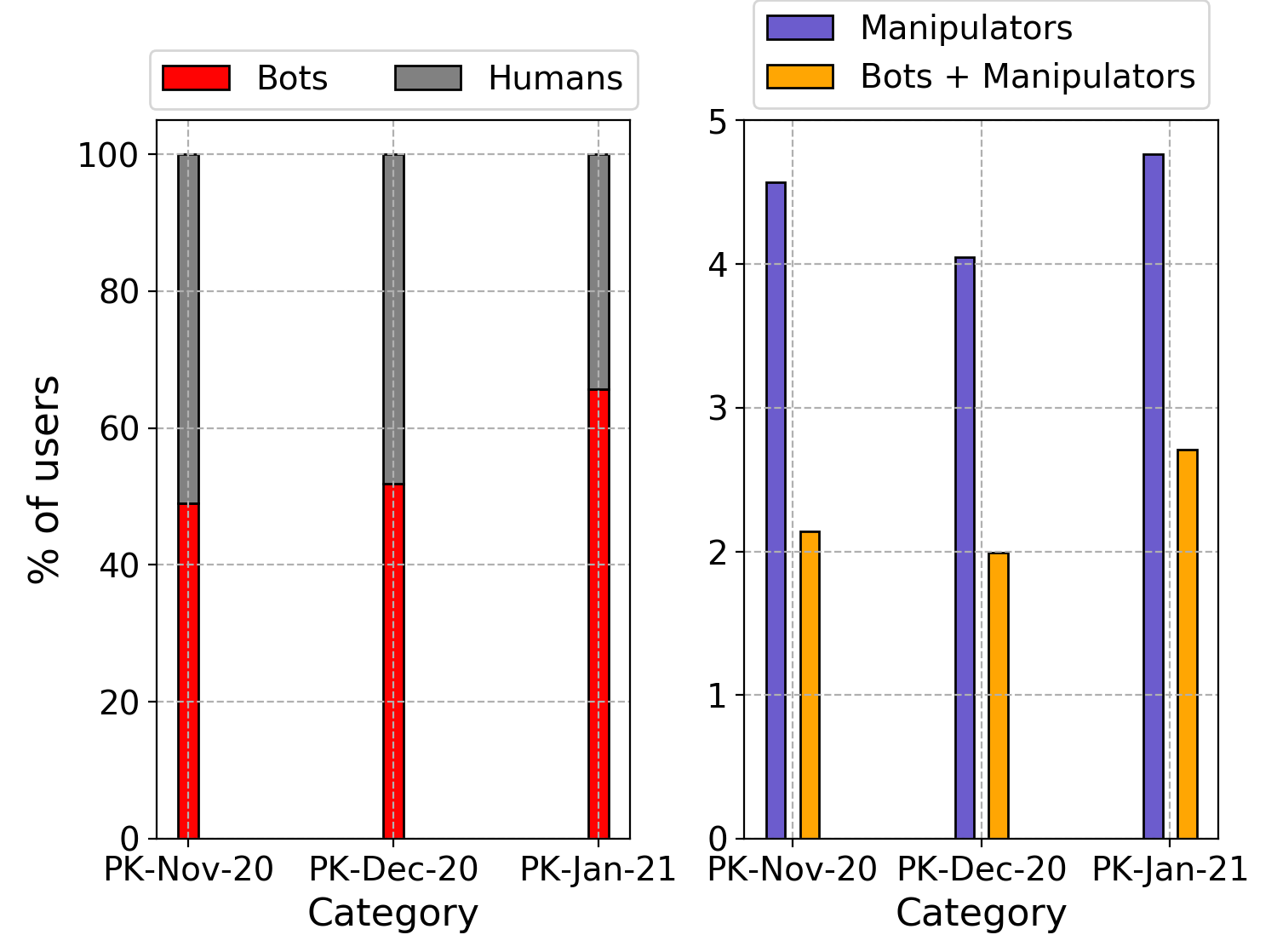}
\caption{Percentage of bots, humans and manipulators -- PK-Trends-Local dataset.} \label{fig:user}
\end{figure}

\section{PK-Trends Analysis} \label{sec:pktrends_analysis}

In this section, first we classify and analyse the content of hashtags in PK-Trends-Local dataset. Next, we identify the malicious users for each hashtag and discuss their distribution in PK-Trends-Local. Finally, we present the category wise analysis of users. 


\begin{figure*}[t] \centering
\includegraphics[width=0.9\textwidth,keepaspectratio]{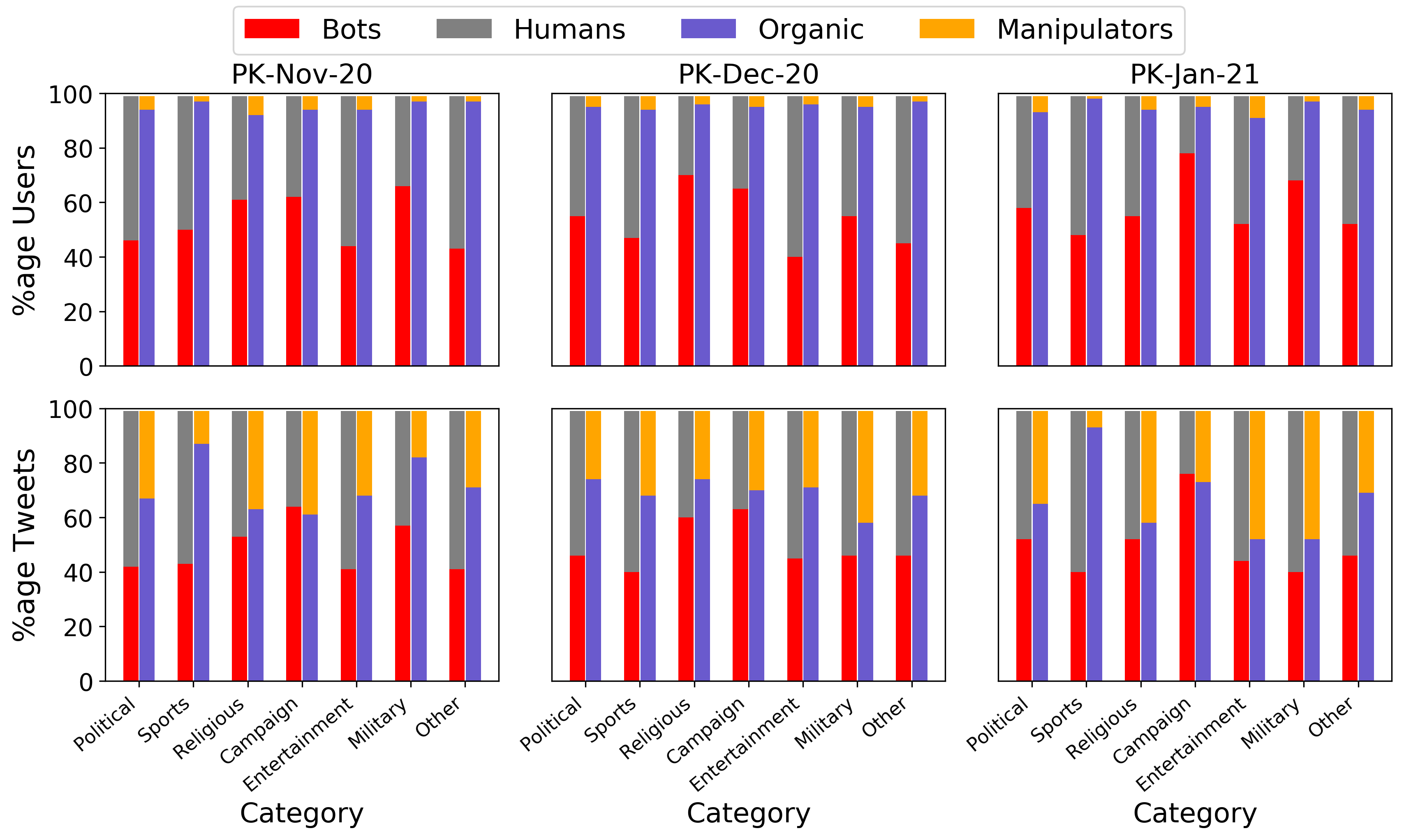} 
\caption{Category-wise percentage of users and tweets of bots, humans and manipulators -- PK-Trends-Local dataset.} \label{fig:bots_dist}
\end{figure*}

\begin{figure*}[t] \centering
\includegraphics[width=1\textwidth]{{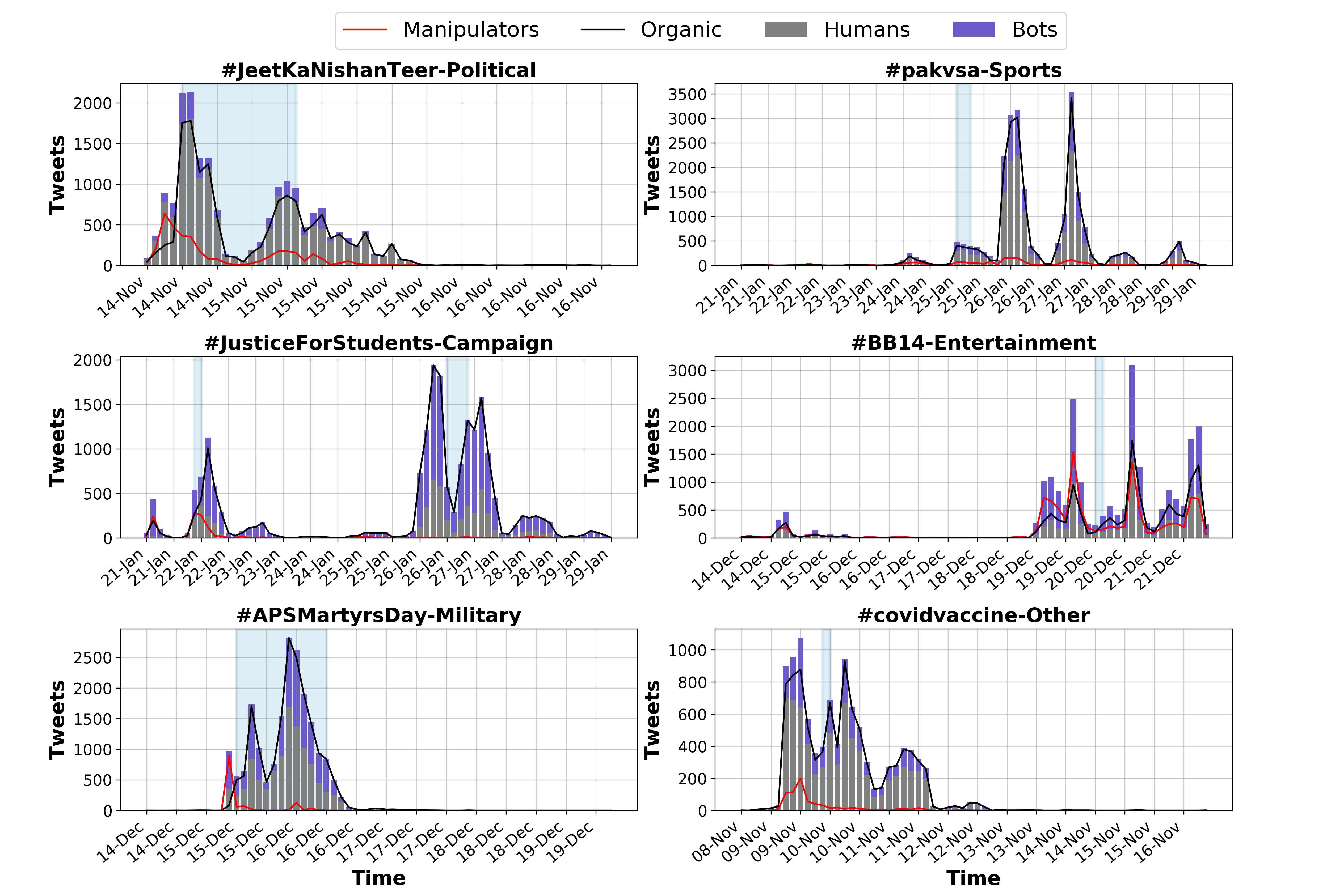}}
\caption{Time series of selected hashtags (highlighted regions represent the time in trending).} \label{fig:time_series}
\end{figure*}

\subsection{Content Analysis}
To begin with the analysis, first, we classify the hashtags in PK-Trends-Local. This is done in order to conduct the user analysis for various hashtag categories.
Figure~\ref{fig:catdist} shows the distribution of hashtags and tweets related to each category in the PK-Trends-Local dataset. The classification results show that $32$-$40$\% hashtags belong to the political category. This result highlights the interest of the general public in politics. In addition, $15$-$32$\% of Twitter trending hashtags belong to the campaign category. Interestingly, $8$-$40$\% tweets belong to this category show the promotional efforts of users for campaign hashtags. Moreover, only $2$-$11$\% hashtags belong to the sports, $8$-$10$\% to the religious, $5$-$15$\% to the entertainment, and $1$-$7$\% to the military category. Also, the `other' category contains less than $15$\% hashtags in three datasets of PK-Trends-Local showing the large coverage of Manipify for analysis trending panel with six pre-determined categories. Zooming into the detailed analysis, Figure~\ref{fig:trends} provides the category-wise distribution of hashtags for each day. We note that the distribution of hashtag categories is intermittent due to the influence of real-world events on the trending panel. For instance, on the anniversary of the shooting at Army Public School (APS) in Pakistan on 16 December, 9 ($35$\%) hashtags related to military class are seen in the trending panel.

Next, we analyze the distribution of natural languages of tweets, hashtag reach, and sentiments for different hashtag categories. As Manipify processes the data for Urdu and English languages only, we inspect the ratio of tweets for English, Urdu, unknown, and other languages as described in Section~\ref{sec:dataset}. Figure~\ref{fig:lang} shows the percentage of tweets of each language belonging to seven categories in PK-Nov-20. We notice that $15$-$75$\% tweets are posted in the English language. Also, the political and religious categories contain $56$\% and $60$\% tweets in the Urdu language, respectively. On the other hand, sports contains $77$\% while the entertainment category contains $50$\% English tweets. Moreover, the dataset contains $60$-$80$\% tweets posted in English and Urdu language showing that the Manipify framework effectively analyzes the predominant part of tweets related to the trending panel. The datasets of PK-Dec-20 and PK-Jan-21 shows a similar pattern for language distribution. From these results, we conclude that the user prefers the local language Urdu to discuss the topics related to political and religious categories. In addition, sports and entertainment categories contain a high percentage of English tweets because such category hashtags are discussed by international users as well.

Figure~\ref{fig:reach} shows the average reach of hashtags related to each category. We observe that the sports category has maximum reach with a limited number of hashtags and tweets as shown in Figure~\ref{fig:catdist}. This result highlights that the substantial reach of the sports category is attributed to the usage of such hashtags by international celebrities. For example, the hashtags \#PakvsSA and \#NZvPAK are used by international cricket players referring to the cricket match of Pakistan versus South Africa and New Zealand versus Pakistan, respectively. Moreover, the religious and campaign category hashtags have a lower value for reach. These results provide an interesting conclusion that religious category hashtags are generally used by normal Twitter users instead of celebrity users. Also, the campaign category hashtags are used for a limited audience from Pakistan.

For sentiment analysis of hashtags in PK-Trends-Local, we manually label the sentiment of all hashtags. Annotators assign the label positive, negative, and neutral by analyzing the text of the hashtag only. After labelling, we notice that $41$\% hashtags have positive while $23$\% express negative sentiment in the dataset. Furthermore, we investigate the category-wise distribution and found that $40$\% of hashtags containing negative sentiment belong to the campaign and $36$\% hashtags belong to the political category. Besides sentiment labelling, the annotators are also asked to label the political hashtags into four classes: 1) faction, 2) personality, 3) slogan, and 4) general discussion. The `faction' class represents the political parties like \textit{\#PTI} while hashtags discussing the political persons such as \textit{\#ImranKhan} are marked as `personalities'. Similarly, slogans of political entities i.e., \textit{\#JeetKaNishanTeer} are organized into `slogans'. Finally, the remaining political hashtags are label as `general' discussion. After labelling, interestingly, all political slogans are labelled as having positive sentiment. In addition, we observe negative sentiment in $20$\%, $37$\%, $26$\% political hashtags related to faction, personalities, and general discussions, respectively. From these results, we draw three interesting conclusions. First, the political and campaign category hashtags are used for negative campaigning and mudslinging as highlighted by a large percentage of negative sentiment hashtags in these categories. Second, hashtags related to political slogans are generated by political factions themselves for their promotion therefore they possess positive sentiments. Finally, hashtags related to factions and personalities are used by political rivals to create political polarization on online social media.

\subsection{User Analysis}
We initiate the user analysis by exploring the behaviour of users in PK-Trends-Local to determine the patterns of manipulation. Figure~\ref{fig:user} shows the distribution of manipulators and organic users in PK-Trends-Local. Interestingly, PK-Nov-20, PK-Dec-20, and PK-Jan-21 contain $4.5$, $4$, and $5$\% manipulators, respectively. Exploring the manipulators further, Figure~\ref{fig:bots_dist} category-wise distribution of manipulators and organic users. Moreover, the percentage of tweets posted by these users is also provided. We observe the presence of only $1-5$\% manipulating users in sports hashtags. This result is anticipated because sports hashtags like \textit{\#PakvsSA} are trended during the real-world event of a cricket match. However, a higher percentage of manipulators is observed in the political and entertainment hashtags with $5$-$10$\% and $8$-$12$\% manipulators, respectively. On the contrary, the religious, campaign, and military categories contain only $2$-$8$\% manipulators. However, focusing on the tweets higher percentages of tweets are seen with a very low percentage of manipulative users. To sum up the results, we conclude that a higher percentage of manipulators in the political hashtags is due to targeted mudslinging generated by the rival political factions~\cite{evans2017mudslinging}. The manipulators generate fake trends of such political hashtags to increase the exposure of their content to Twitter users. In addition, the trends of entertainment hashtags are generated with pre-planned coordinated efforts to promote TV shows, movies and music. 

Figure~\ref{fig:user} shows that PK-Nov-20, PK-Dec-20, and PK-Jan-21 contain $50$, $52$, and $64$\% bot accounts. In addition, the bot accounts are highly suspected to play a key role in the manipulation of the trending hashtags~\cite{Howmucht23:online}. Therefore, we further investigate the users identified as bots as well as manipulators. Figure~\ref{fig:user} shows the percentage of such users. We observe that $2.1$\%, $2.0$\%, $2.7$\% accounts are identified as bot as well as manipulators in PK-Nov-20, PK-Dec-20, and PK-Jan-21, respectively. Interestingly, the percentage of bot users involved in manipulation is consistent in all data points. Whereas, the percentage of bots only is least in PK-Nov-20 and highest in PK-Jan-21. This result is expected because the accounts generating automated activity are suspended by Twitter~\cite{Manipula68:online}. Considering that the data related to all three datasets in fetched in February 2021, the tweets posted by such deactivated or suspended accounts are not fetched in PK-Trends-Local.

Furthermore, Figure~\ref{fig:bots_dist} also shows the category-wise percentage distribution of bots and humans users along with tweets posted by these users. We notice that the campaign category has the largest percentage of bots with $60$-$78$\% bots. Similarly, the sports hashtags contain $45$-$50$\% bots. In addition, $42$-$50$\% tweets related to political hashtags are created by bot users. From these results, we conclude that bot accounts are used for the promotion of campaign and entertainment hashtag~\cite{gilani2016stweeler}. Moreover, sports hashtags like \textit{\#PakvsSA} are used by bot accounts to provide live updates related to match. Interestingly, for political hashtags, the bot accounts are used for promotion as well as to provide live updates related to political activity. For instance, the political hashtag of \textit{\#JeetKaNishanTeer} is promoted using bot accounts with political motives while the hashtag of \textit{\#GBElection2020} is used to provide live updates regarding by-election.

\begin{table*}[t] \centering
\caption{Hashtag pairs generated by political rivals on Twitter} \label{tab:pairs}
\renewcommand{\arraystretch}{1.3}%
\begin{tabular}{|l|c|c|c|c|c|c|c|c|}\hline & & 
& \multicolumn{2}{c|}{\textbf{Bots}} 
& \multicolumn{2}{c|}{\textbf{Manipulators}} 
& \multicolumn{2}{c|}{\textbf{Bot Manipulators}} \\ \hline
\textbf{Pair}    & \textbf{Hashtag}  & \textbf{Trend Time} &
\textbf{\%Users} & \textbf{\%Tweets} &
\textbf{\%Users} & \textbf{\%Tweets} & 
\textbf{\%Users} & \textbf{\%Tweets} \\ \hline\hline
\textbf{1-ORIG} & \includegraphics[width=0.2\textwidth,keepaspectratio]{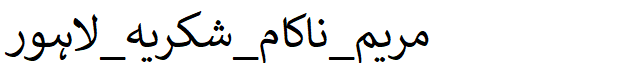} 
& 2020-12-14 06:00 & 58 & 52 & 1 & 8 & 1 & 2 \\ \hline
\textbf{1-RESP} & \includegraphics[width=0.2\textwidth,keepaspectratio]{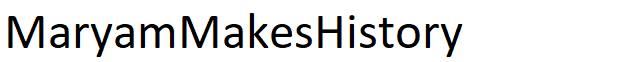} 
& 2020-12-14 10:00 & 52 & 47 & 1 & 9 & 0.4 & 2 \\ \hline
\textbf{2-ORIG} & \includegraphics[width=0.2\textwidth,keepaspectratio]{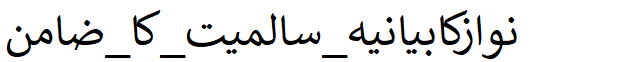} 
& 2020-12-20 14:00 & 52 & 42 & 10 & 38 & 4 & 15 \\ \hline
\textbf{2-RESP} & \includegraphics[width=0.2\textwidth,keepaspectratio]{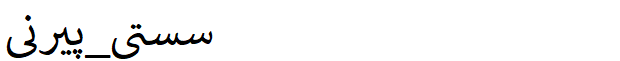} 
& 2020-12-20 18:00 & 56 & 57 & 11 & 51 & 5 & 29 \\ \hline
\textbf{3-ORIG} & \includegraphics[width=0.2\textwidth,keepaspectratio]{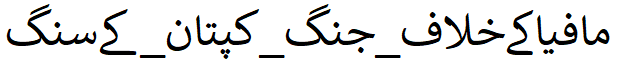} 
& 2021-01-28 13:00 & 48 & 46 & 9 & 39 & 3 & 17 \\ \hline
\textbf{3-RESP} & \includegraphics[width=0.2\textwidth,keepaspectratio]{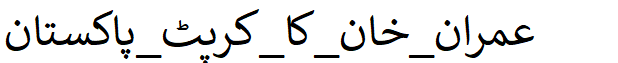} 
& 2021-01-28 19:00 & 56 & 48 & 9 & 55 & 4 & 24 \\ \hline
\textbf{4-ORIG} & \includegraphics[width=0.2\textwidth,keepaspectratio]{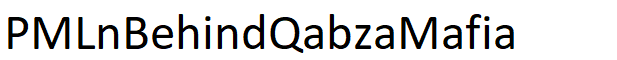} 
& 2021-01-24 13:00 & 53 & 47 & 5 & 22 & 2 & 10 \\ \hline
\textbf{4-RESP} & \includegraphics[width=0.2\textwidth,keepaspectratio]{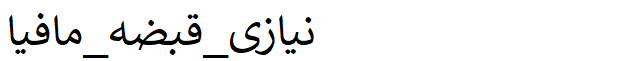} 
& 2021-01-24 17:00 & 56 & 55 & 4 & 25 & 2 & 17 \\ \hline
\end{tabular}
\end{table*}

To provide an in-depth analysis of PK-Trends-Local, Figure~\ref{fig:time_series} shows the time series plot of six hashtags related to political, sports, campaign, entertainment, military, and `Other' category. In particular, the time-series plot of the number of tweets posted by bots and humans as well as a manipulator and organic users are presented. While the shaded region highlights the time a hashtag is part of the trending panel. First, focusing on manipulators, we observe that the political hashtag of \textit{\#JeetKaNishanTeer} contains more tweets posted by manipulators before trend time. While the organic users discuss the hashtag after the trending time highlighting that manipulators limit their activity after making the hashtags in the trending panel. The military hashtag of \textit{\#APSMartyrsDay} shows a similar pattern. Similarly, the entertainment hashtag \textit{\#BB14} has a higher number of tweets by manipulators before the hashtag is seen in the trending panel. In addition, the manipulators remain active after the hashtag is seen in the trending panel. For the entertainment category, this result highlights that manipulators not only trend the hashtags but also put effort to disseminate content related to hashtag. However, this observation does not hold for the sports hashtags. The sports hashtag \textit{\#pakvsa} contain very few tweets from manipulators. Similarly, the limited number of tweets related \textit{\#covidvaccine} and \textit{\#JusticeForstudents} hashtags are posted by manipulators. This lower percentage of tweets generated by manipulators shows that these topics are actually discussed by organic Twitter users.

We further analyse the time-series plot for the tweets created by the bot and human users. For example, a time-series plot for the hashtag of \textit{\#covidvaccine} from the `Other' category shows that this trend is largely used by human accounts. However, bot accounts have little participation in generating the content for this hashtag. In addition, the sports hashtag \textit{\#pakvsa} is used by bots after being initiated by human users. Similarly, the hashtag \textit{\#JeetKaNishanTeer} from the political category has a small number of bot accounts. In addition, the time-series of campaign hashtag \textit{\#JusticeForStudents} shows different behaviour. The content of this hashtag is majorly posted by bot accounts and later humans joined the discussion after seeing it in the trending panel. The hashtag \textit{\#APSMartyrsDay} from the military category have a longer time-span (24 hours) in the trending panel and the $10$-$50$\% tweets posted by bot users. Finally, the entertainment hashtag \textit{\#BB14} shows a large number of tweets by bots. Interestingly, this hashtag also contains a large number of tweets generated by manipulators. From this result we conclude the manipulation is performed by bot accounts for the entertainment category.

From the analysis of sample hashtags, we conclude three observations. First, the behaviour of manipulator and bot accounts vary for each category, emphasizing that generalized patterns cannot be observed among different categories. Second, the entertainment and campaign hashtags rely on bot accounts for the manipulation while political hashtags are manipulated by human users. Finally, for entertainment hashtags, manipulators not only manipulate the trend to the trending panel but also continue generating content to keep users engaged.

\section{Case Study} \label{sec:casestudy}
Figure~\ref{fig: architecture} shows the applications of our framework. In this regard, this section presents two real-world case studies to show the efficacy of Manipify. First, we identify and analyze hashtag-wars in trending panel of Pakistan using Pk-Trends datasets. Next, we examine the fake trend spreading anti-state propaganda on Twitter.

\begin{figure*}[t] \centering
\includegraphics[width=0.76\textwidth]{{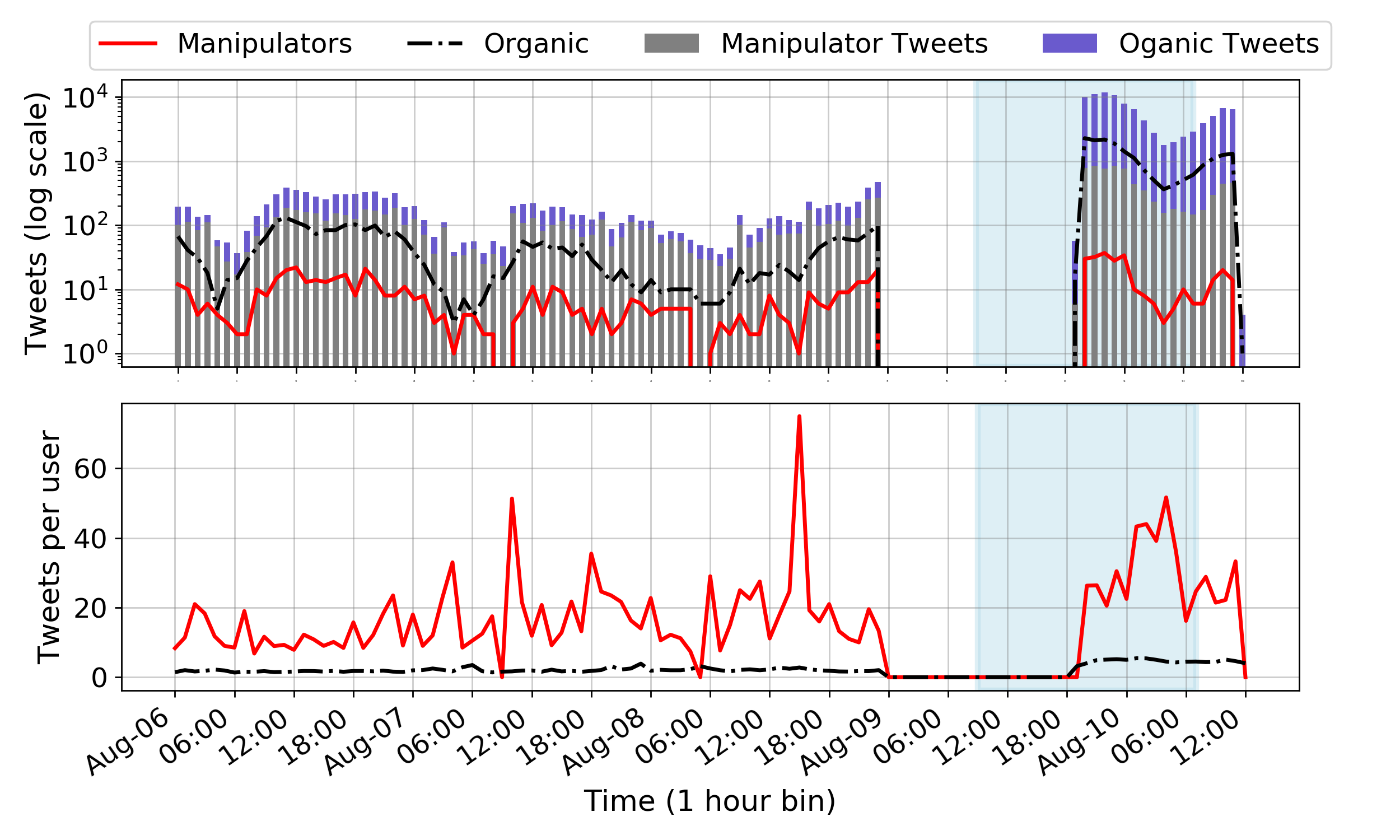}}
\caption{Time series of \#SanctionPakistan (Trending period is highlighted).} \label{fig:sanction}
\end{figure*}

\subsection{Hashtag wars}

Hashtag wars is phenomenon of disseminating the hegemonic narrative on social media platform to sabotage the visibility of conversations with opposing views. In this regard, we manually analyze trending hashtags in Pk-Trends dataset and notice that few hashtags trending at the same time contain the opposing narrative of rival parties. Table~\ref{tab:pairs} shows four pairs of trending hashtags seen at same time in the trending panel. Each pair contains hashtags with opposite narrative related two major political parties of Pakistan Tehreek Insaf (PTI) and Pakistan Muslim league (PMLN). Moreover, hashtag that appears first in the trending panel is the original (ORIG) hashtag and that appears later is the response (RESP) hashtag. We initiate the analysis of these hashtags by comparing the trending time of `ORIG' hashtag with the time when the first tweet of `RESP' Hashtag is posted. We notice that first tweet of `RESP'  hashtag is posted after the hashtag `ORIG' is seen in the trending panel for each pair. This analysis depicts that the trend of `RESP' Hashtag is deliberately created by manipulation with the aim of spreading opposing political narrative. In addition to manipulators, we also identify three types of highly active users posting content related to both hashtags in hashtag-war. First, few users belong to online trend services and post all current trends periodically in their tweets. Second, we discover news channels and journalist's accounts that actively post content related to trending hashtags. Finally, users also post memes using both trending hashtags for maximum visibility and reach.

\subsection{Anti-state Propaganda}

Recently, manipulators squads are identified creating fake trends to malign the reputation of different countries~\cite{Golovchenko:StatePropaganda}. The identification of such users is a prime real-world application of our research. In this regard, we notice that trend of a hashtag \#SanctionPakistan is created to damage the repute of Pakistan worldwide. This trend also caught the attention of the Pakistani government and a detailed analysis report is issued by the government highlighting the key factors in creating this trend~\cite{Analysis:Sanction}. This provides us with an excellent opportunity to compare the results of our user classifier module with statistics shared in the government report. According to the report, coordinated efforts is made to spread anti-Pakistan propaganda using this hashtag~\cite{sanctionpak:online}.

This hashtag trended in Pakistan on \textit{Aug 9, 2021}. We fetch all tweets for \#SanctionPakistan using Twint~\cite{twint} on \textit{Aug 12, 2021} posted before the date of data collection. We collected a total of 113,855 original tweets posted by 23,012 unique users related to this hashtag. Using our manipulator classifier, we identify $800$ ($3.5$\%) users involved in manipulation. These users posted $17,425$ ($15.3$\%) tweets. Figure~\ref{fig:sanction} shows the distribution of manipulators and organic users along with tweets posted by these users with a time bin of one hour. We also plot the average number of tweets per user ($Tweets$\textsubscript{user}) for each hour. Moreover, the trending period of the hashtag is highlighted as a shaded area. We observe that the volume of tweets by organic users increases enormously after the trending time of hashtags. Moreover, we notice that $Tweets$\textsubscript{user} for manipulators lie in the range of $7$ to $75$. While, for the organic users $Tweets$\textsubscript{user} vary from $1$ to $5$. Surprisingly, we observe 35-468 tweets per hour before the hashtag is seen in the trending panel. This observation is imperceptible because manipulators post a large number of tweets before the trending time. For in-depth analysis, we manually analyze and compare the tweets of manipulators identified in report~\cite{sanctionpak:online}. In the report, only the top 9 contributors for \#SanctionPakistan hashtag are presented on the basis of total tweets and retweet count. We observe that 2 users have been removed from Twitter. Therefore, our collected data of 113K tweets do not contain the tweets of removed users. This result provides two insightful findings. First, Manipify needs complete data for precise performance which can be obtained by collecting tweets in real-time. Passive collection of data results in loss of valuable data. Second, deletion of tweets is an important feature that can be included in Manipify to enhance its performance. However, in presence of complete data, we believe that Manipify can automatically detect the large potion of manipulators using user behaviour and content features. In contrast, manual analysis relying only on the number of tweets/retweets is a time-consuming and inefficient approach to detect all users manipulating the trend. Moreover, Manipify can be adapted to detect manipulators posting content in any natural language.

\section{Conclusion} \label{sec:conclusion}

The dynamic behaviour of manipulators on the Twitter platform makes the automatic detection of such users a challenging task. This challenge is further exacerbated due to the involvement of both human and bot accounts in manipulation. In this paper, we identify and study the characteristics of users manipulating the trending panel of Pakistani Twitter. For this purpose, we propose a novel framework of `Manipify' to detect and analyze the manipulators with $0.91$ accuracy. We further identify bot accounts and notice that human accounts are more involved in manipulation. Manipify also classifies the hashtags into six categories to analyze the behaviour of users across different categories. Testing of the framework on our Pk-Trends dataset highlights that political and entertainment category hashtags are the most manipulated trends. Also, we find $4.4$\% of user accounts as manipulators generating $25.6$\% of tweets. This shows that a larger percentage of content is posted by a small number of manipulators. Finally, we evident the significance of the research by presenting two real-world case studies of hashtag-wars and anti-state propaganda.   

In future, we plan to extend the framework to identify users who spread hate speech and propaganda in a coordinated manner. Besides, considering the multi-faceted 5th generation war on social media, we will work on location identification of manipulators working for rival countries to create polarization in the society.

\section{Acknowledgement}
\label{sec:acknowledgement}
This research work was funded by Higher Education Commission (HEC) Pakistan and Ministry of Planning Development and Reforms under National Center in Big Data and Cloud Computing.

\bibliographystyle{IEEEtran}
\bibliography{refer}

\end{document}